# One-reactor vacuum and plasma synthesis of transparent conducting oxide nanotubes and nanotrees: from single wire conductivity to ultra-broadband perfect absorbers in the NIR

Javier Castillo-Seoane,[a,b] Jorge Gil-Rostra,*[a] Víctor López-Flores,[a] Gabriel Lozano,[c] F. Javier Ferrer,[d] Juan P. Espinós,[a] Kostya (Ken) Ostrikov,[e,f] Francisco Yubero,[a] Agustín R. González-Elipe,[a] Ángel Barranco,[a] Juan R. Sánchez-Valencia,[a,b] Ana Borrás*[a]

The eventual exploitation of one-dimensional nanomaterials yet needs the development of scalable, high yield, homogeneous and environmentally friendly methods able to meet the requirements for the fabrication of under design functional nanomaterials. In this article, we demonstrate a vacuum and plasma one-reactor approach for the synthesis of the fundamental common element in solar energy and optoelectronics, i.e. the transparent conducting electrode but in the form of nanotubes and nanotrees architectures. Although the process is generic and can be used for a variety of TCOs and wide-bandgap semiconductors, we focus herein on Indium Doped Tin oxide (ITO) as the most extended in the previous applications. This protocol combines widely applied deposition techniques such as thermal evaporation for the formation of organic nanowires serving as 1D and 3D soft templates, deposition of polycrystalline layers by magnetron sputtering, and removal of the template by simply annealing under mild vacuum conditions. The process variables are tuned to control the stoichiometry, morphology, and alignment of the ITO nanotubes and nanotrees. Four-probe characterization reveals the improved lateral connectivity of the ITO nanotrees and applied on individual nanotubes shows resistivities as low as $3.5 \pm 0.9 \times 10^{-4}$ $\Omega \cdot cm$, a value comparable to single-crystalline counterparts. The assessment of diffuse reflectance and transmittance in the UV-VIS range confirms the viability of the supported ITO nanotubes as a random optical media working as strong scattering layers. Further ability to form ITO nanotrees opens the path for practical applications as ultra-broadband absorbers in the NIR. The demonstrated low resistivity and optical properties of these ITO nanostructures open the way for their use in LEDs, IR shield, energy harvesting, nanosensors, and photoelectrochemical applications.

## Introduction

Transparent electrodes are ubiquitous in smartphones, touch display panels, portable tablets, nanostructured solar cells, micro-energy harvesters, transparent heaters, and other advanced nanotechnology enhanced devices. The transparent conducting layers are used in optoelectronics, either as collectors of the charges formed at the photoabsorbers, e.g. in photovoltaic devices such as perovskite, dye-sensitized, and hybrid solar cells or as functional layers that supply charge carriers, as in light-emitting diode displays (LEDs).[1–3] The most commonly used in industry transparent electrode is made of indium-doped tin oxide (ITO), a transparent conducting oxide (TCO) showing both high transparency (>90%) and high electrical conductivity (10 $\Omega$ sq$^{-1}$ on glass).[4] However, several shortcomings such as indium scarcity and ITO inherent brittleness, which reduce its compatibility with roll-to-roll processing and flexible plastic substrates, have stimulated intense research into alternative indium-free oxides[5,6] and flexible nanomaterials like metal networks, carbon nanotubes, or graphene.[7] Nevertheless, the application of ITO thin films remains dominant in the commercial market due to their outstanding performance and mature manufacturing technologies which mostly rely on various vapor-phase processes.[8]

The ITO transparent electrodes are commonly utilized in the form of polycrystalline thin films. Recent advances in the synthesis and device integration of nanomaterials have stimulated the rapid development of low-dimensional TCOs as nanoparticles, nanocolumns, nanowires and nanotubes, as well as micro/nano-patterned surfaces.[9–14] One-dimensional (1D) TCO nanomaterials such as nanowires (NWs) and nanotubes (NTs) are excellent candidates for the assembly of photodetectors,[15,16] electron emitters,[17] phototransistors,[10] light-emitting diodes (LEDs),[18] biological and chemical sensors,[19,20] UV nanolasers, anti-reflective coatings and broadband saturable absorbers for solid-state lasers.[21] The most widespread methods for the preparation of ITO 1D nanostructures include catalytic or self-catalytic growth of single crystalline nanowires[22,23] by vapor–solid (VS) and vapor-liquid-solid VLS growth mechanisms[24] and epitaxial growth[17,25] yielding single-crystalline nanostructures. Besides, several template-assisted fabrication protocols have been reported as the use of hard (carbon nanotubes (CNTs),[26] anodized aluminum oxide (AAO)[27,28]) and soft (polystyrene,[29,30] polycarbonate[31] membranes) templates. Indeed, 1D TCO nanostructures could help to overcome the low flexibility of the ITO thin films.[8,14] However, two challenges remain unresolved, namely the fabrication of nanostructured electrodes with laterally connected nanostructures and the development of scalable deposition methods compatible with a variety of substrates.

Consequently, the aim of this article is twofold. On one hand, we will demonstrate the one-reactor vacuum and plasma deposition method for the formation of 1D and 3D supported ITO nanotubes. On the other hand, we will characterize these nanostructures focusing on their fundamental optical and electronic properties of direct relevance to their applications in optoelectronic devices. We further reveal the possibility of lithography-free nanostructure arrangements suitable for the fabrication of next-generation broadband absorbers in the NIR.

The approach, outcomes, and potential applications of the results of this work are presented in **Scheme 1**. The first objective of this work is to develop an advanced soft-template method and a full vacuum "one-reactor" configuration to fabricate ITO 1D (nanotubes) and 3D

(nanotrees) supported on processable substrates, under mild temperature conditions that might outperform the available state-of-the-art. The soft template methods are based on the use of template materials with a flexible structure such as organic molecules, polymers, or biopolymers.[11,32] These approaches show advantages like the use of low-cost template-materials, good repeatability, simplicity, and easy removal of the template.[33] The key innovation herein relies on the use of self-assembled supported single-crystalline organic nanowires (ONWs) as 1D or 3D scaffolds. So far, these small-molecule ONWs have been exploited as active components in organic electronics, optoelectronic devices, and nanosensors.[34,35] Recently, the application of such low-cost nanomaterials has been reported for the formation of core@(multi)shell NWs, NTs, and 3D counterparts as well as precursors for porous metal and metal oxide thin films.[36–41] Herein, we implement, for the first time, the fabrication of the ITO supported nanostructures by combining well-established and scalable industry-relevant nanofabrication processes. Namely, our system integrates the proprietary vacuum and plasma processes carried in the same reactor, under mild vacuum conditions.

As demonstrated below, our new proprietary approach allows to:

i) Integrate sequential and/or simultaneous fabrication/processing steps into the same reactor to reduce the number of vacuum chambers and minimize sample transfers between them (thus reducing hardware cost, processing time, and energy consumed in the process);

ii) Produce every layer of the active system by a vacuum procedure (avoiding liquid solvents and reducing waste by-products, increasing reproducibility and interface stability);

iii) Generate microstructure-tailored nanoarchitectures with excellent control of the interfacial composition of sequentially deposited materials with either sharp or gradual/graded interfaces including organic, polymeric, inorganic, and hybrid compositions.

Remarkably, our approach surpasses the most common TCO magnetron sputtering deposition by the unprecedented production of polycrystalline ITO nanotubes and nanotrees. These nanostructures feature tuneable thickness and length, excellent optoelectronic properties, and low resistivity, which collectively make them promising candidates for optical random media in the visible (VIS) and near-perfect absorbers in the Near Infrared (NIR) ranges.

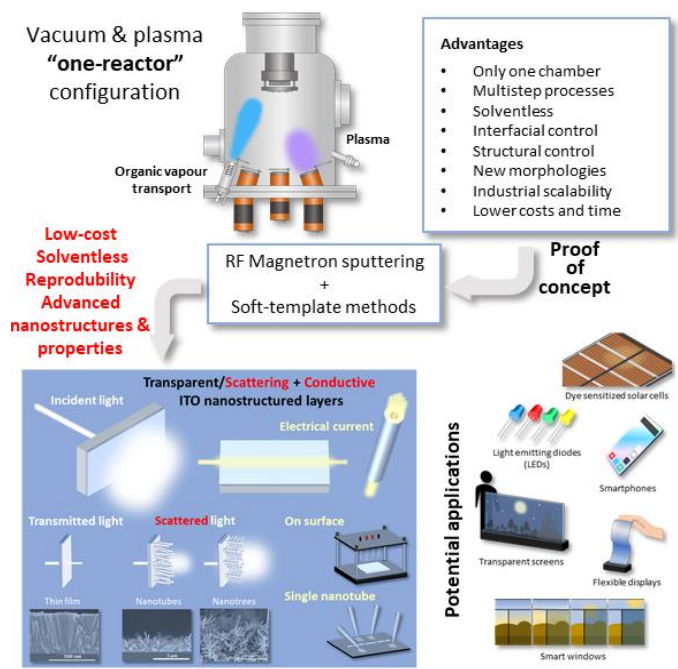

**Scheme 1.** The one-reactor approach for scalable vacuum and plasma deposition of nanostructured TCOs for diverse applications. The schematic includes advantages of the procedure for the development of 1D and 3D nanotubes, the proof-of-concept carried out with a transparent conducting oxide, concretely ITO, and several examples of the foreseeable applications of such nanomaterials.

## Results and discussion

**One-reactor fabrication of ITO nanotubes and nanotrees. Microstructural and chemical characterization.**

We have fabricated the ITO layers and shells under two different working pressure conditions (see detailed information in the Experimental Section). Thus, the acronym LP refers to the "low-pressure conditions" during the magnetron sputtering of the ITO, i.e. $5,0\cdot 10^{-3}$ mbar; meanwhile, HP labels the "high-pressure conditions", i.e. $2,0\cdot 10^{-2}$ mbar. It is worth stressing though that both LP and HP can be considered mild vacuum conditions. The labels also inform on the morphology of the samples, differentiating between thin films (TF), nanotubes (NT), and nanotrees (NTrees). Figures 1 and 2 gather representative SEM images of the step-by-step process for the formation of the nanostructures.

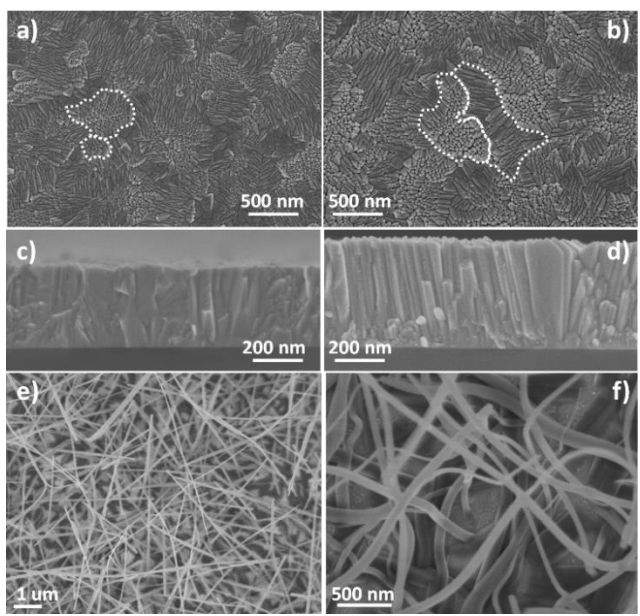

**Figure 1. Steps 1 and 2 of the ITO nanostructure synthesis process.** Representative SEM images including LP and HP thin films (a, c) and (b, d) correspondingly and ONWs formation on the surface of the LP ITO film (e, f). Panels c) and d) show the cross-section images of the thin films. The dashed lines are included to guide the eyes on the different sizes of the crystalline domains.

See Scheme S1 in the Supplementary Material Section for a pictorial representation of the protocol. Figure 1 a) - d) compares the normal and cross-section views of the thin films prepared under the two Ar pressure conditions, LP (a,c) and HP (b,d). At first glance, they present the same morphology with triangular and elongated features coexisting on the surface but forming patches or domains. These domains contain motives with similar morphology and orientation. Close inspection of the planar view images reveals that the size of these domains is larger for the HP conditions, i.e. we count a smaller number of differentiated domains for the same area as highlighted by the dashed lines bounding the crystalline domains for the two samples.

The cross-section views show two different types of morphologies with a characteristic columnar formation. Samples deposited under HP conditions present a better defined columnar cross-section (see panel c) than the LP films. The comparison of these two images also reveals that the growth rate for the HP condition is higher compared to the LP case, with a growth rate of $5.3 \pm 0.1$ nm s$^{-1}$ and $4.4 \pm 0.2$ nm s$^{-1}$, respectively. In the following experiments, we use this finding to deposit ITO films and shells with comparable final thicknesses for both HP and LP conditions.

The following step consists of the formation of phthalocyanine organic nanowires (ONWs) on an ITO thin film acting as the nucleation surface. For such purpose, in Step 1 an ITO thin film is deposited under LP conditions with a thickness of ~120 nm to form the "seed" growth surface. Such a thin ITO layer improves the conductivity of the system while being compatible with flexible substrates. In Step 2, the formation of the ONWs is carried out in the same reactor by placing a low-temperature evaporator facing towards the substrates (see Experimental Section). After several preliminary experiments, the pressure of Ar and temperature of the substrates were established at $1 \cdot 10^{-3}$ mbar and 210 ºC to form a high density of supported ONWs as shown in Figure 1 e-f). The nanowires mostly present a square footprint with the mean width and length of 100 ± 30 nm and 4.0 ± 1.6 µm, respectively (see Figure S1 and Table S1 for the statistics carried out by image analysis). The estimated density of ONWs is 6.5 ONWs µm$^{-2}$. Such formation of ONWs on the surface of the ITO thin films deposited in the same reactor is the first step towards the demonstration of the "one-reactor" approach.

It is important to stress herein that by this procedure the substrate is not limited to an ITO thin film or surface. Thus, we have previously demonstrated the compatibility of this soft-template formation with a variety of supports, including metal and metal oxide layers and nanoparticles, polymers, and organic systems. This is also extended, for instance, to the direct formation of the ONWs on device architectures and pre-formed electrodes.[38,42–45]

In Step 3, the ONWs are coated with ITO shells at both LP and HP conditions yielding in both cases a mean thickness of the NTs about 200 nm for both conditions. The panels a) to f) in Figure 2 show the SEM images at different magnifications (see also Figure S1 and Table S1). From Figure 2 b and e), it is easy to see a different growth for the two conditions. On one hand, LP ITO nanotubes show a homogenous morphology along their entire structure with non-defined domains. On the other hand, HP ITO nanotubes exhibit well-defined morphological domains, as in extension of such characteristics of the corresponding thin films microstructure. Although the alignment is not as pronounced as that found with the soft-template approach using plasma-enhanced chemical vapour deposition,[36,50] the cross-sectional images suggest that the ITO nanotubes (Figure 2 c and f) present a slight vertical alignment in comparison with the organic nanowires (Figure 1 e and f).

In the final step, the leftovers of the organic nanowires remaining after the ITO deposition are easily removed by annealing at 350ºC under air or vacuum conditions (within the same reactor adding oxygen). The organic molecules evacuate the core through the pores of the ITO shell and substrate leaving no traces as we have previously shown by HAADF-STEM on ZnO and $TiO_2$ nanotubes.[45] This straightforward removal of the template is a key advantage of the method and will be also supported by the optical properties discussed below. Finally, the samples were annealed at 350 ºC in Ar to ensure fine transparency (see Experimental section).

Pursuing the sensing, photoelectrochemical and electrocatalytic applications of TCOs, it is imperative to generate highly interconnected, 3D arrangement and open nanostructures yielding high-surface-area electrodes. Hence, the next step was to fabricate three-dimensional (3D) ITO nanotubes or nanotrees (NTrees). For such purpose, steps 2 and 3 are repeated using the as-grown ITO nanotubes as a substrate (see Schematic S1). Figure 2 g-l) presents SEM images of these nanostructures. In both cases, LP (g-i) and HP (j-l), the morphology resembles a high density of branched nanotubes or nanotrees with a "trunk" surrounded by "branches" distributed on the lateral surfaces of the "trunks" (Figure 2 i) and l)). If we compare both nanostructures, it is obvious that LP nanotrees lead to longer sizes (Figure 2 g) and h)), whereas for HP samples the branches are well distributed over all the substrate leading to smaller sizes (Figure 2 j) and k)) and a higher density than the LP sample. These differences arise from the diverse roughness and motives on the surface of the primary nanotubes yielding such slightly different formation of organic nanowires.

With the extension of the protocol from 1D to 3D by simple iteration of the process, we further demonstrated that the one-reactor approach provides a straightforward way for the fabrication of complex systems by combining two different deposition techniques in the same reactor, i.e. the thermal evaporation of organic molecules and magnetron sputtering of a transparent conducting oxide. On the other hand, we proved that the soft-template procedure based on the use of single-crystalline ONWs as a supported 1D scaffold can be extended to the fabrication of shells by magnetron sputtering, one of the most industrially spread vacuum deposition processes.

TEM images in Figure 3 of individual nanotubes allow the detailed comparison of the morphology and crystalline structure for LP and HP deposition conditions. The lighter area along the axis of the nanotubes corresponds to the cavity left after the organic template sublimation. It is also possible to confirm the growth of columnar and triangular features from such axis forming a conformal shell or wall that are characteristic to the nanotube morphology. The HP nanotubes show larger crystals (65 ± 6 nm) than LP ones (46 ± 6 nm), considering the grains grow outwards from the template hole. The HRTEM images (Figure 3 e) and f)) provide information of the crystalline planes revealing multiple interplanar distances for the LP and HP conditions. The prevalent distances of 5.15 Å and 2.68 Å are related to the (2 0 0) and (4 0 0) planes for LP nanotubes, respectively. The distances of 4.32 Å and 3.05 Å for HP conditions correspond to (2 1 1) and (2 2 2) planes, respectively. These results are in a good agreement with the XRD patterns for the reference thin films, supported nanotubes, and nanotrees presented in Figure 3 g). Notably, the $In_2O_3$ crystalline structure was reproduced in all of them. We emphasize that the (2 1 1), (2 2 2), and (4 0 0) planes show the highest intensity. The crystal size (Figure S2) is generally lower than 100 nm as determined using the Scherrer approximation for all families of planes and under both deposition conditions. However, planes such as (2 1 1) in the HP thin film sample or (2 2 2) in LP nanotubes sample present larger sizes (168.5 and 178.7 nm, respectively). ITO nanotrees prepared at HP conditions display the same characteristic planes as the nanotube counterparts, indicating that both trunks and branches feature the same microstructure.

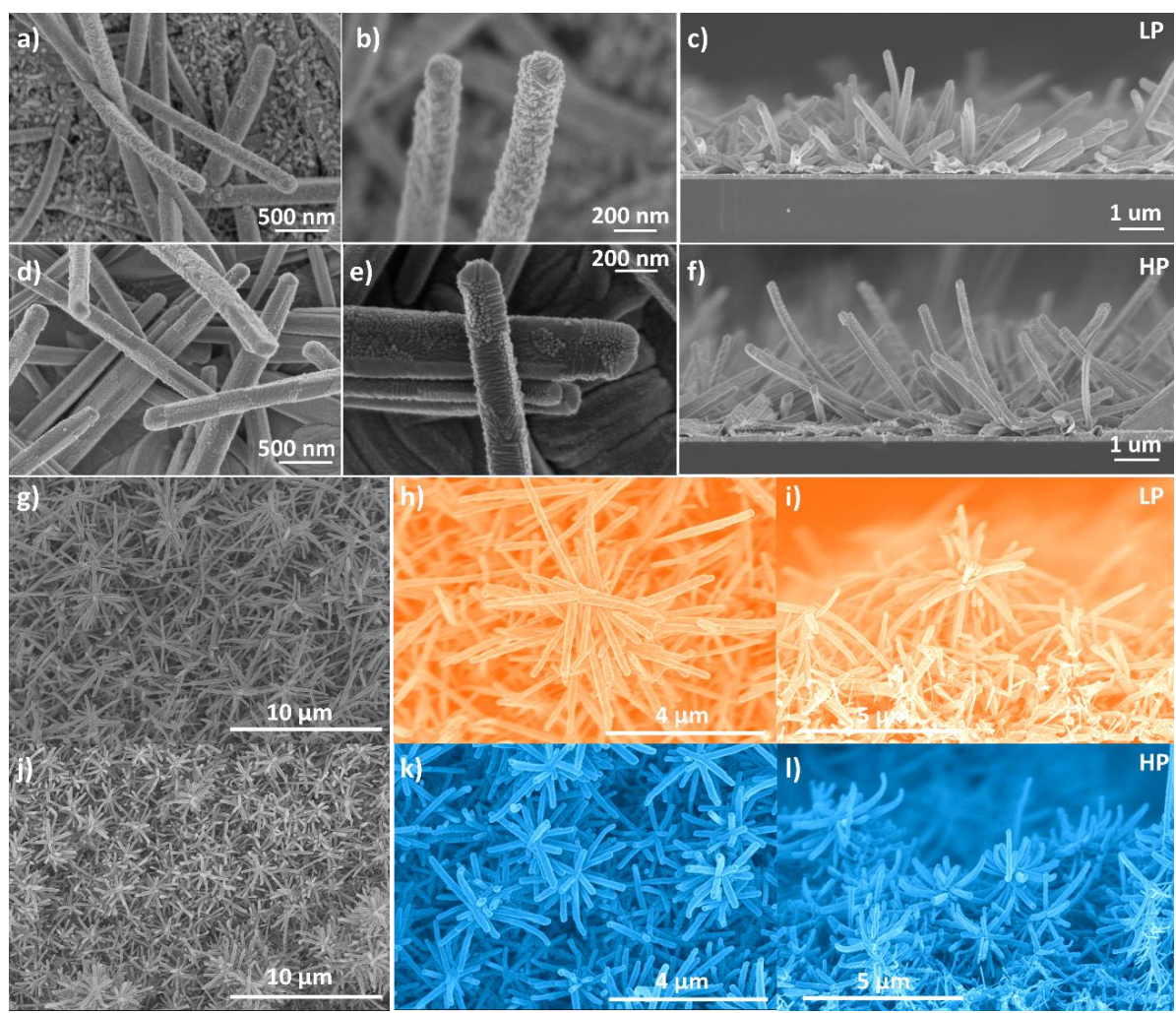

**Figure 2. Supported ITO nanotubes and nanotrees manufactured by the scalable one-reactor approach.** Representative SEM images including nanotubes (a-f) and nanotrees (g-l), with micrographs a-c) and g-i) corresponding to samples fabricated under LP conditions and d-f) and j-l) under HP conditions. Panels on the right-side display cross-section characteristic views of the samples.

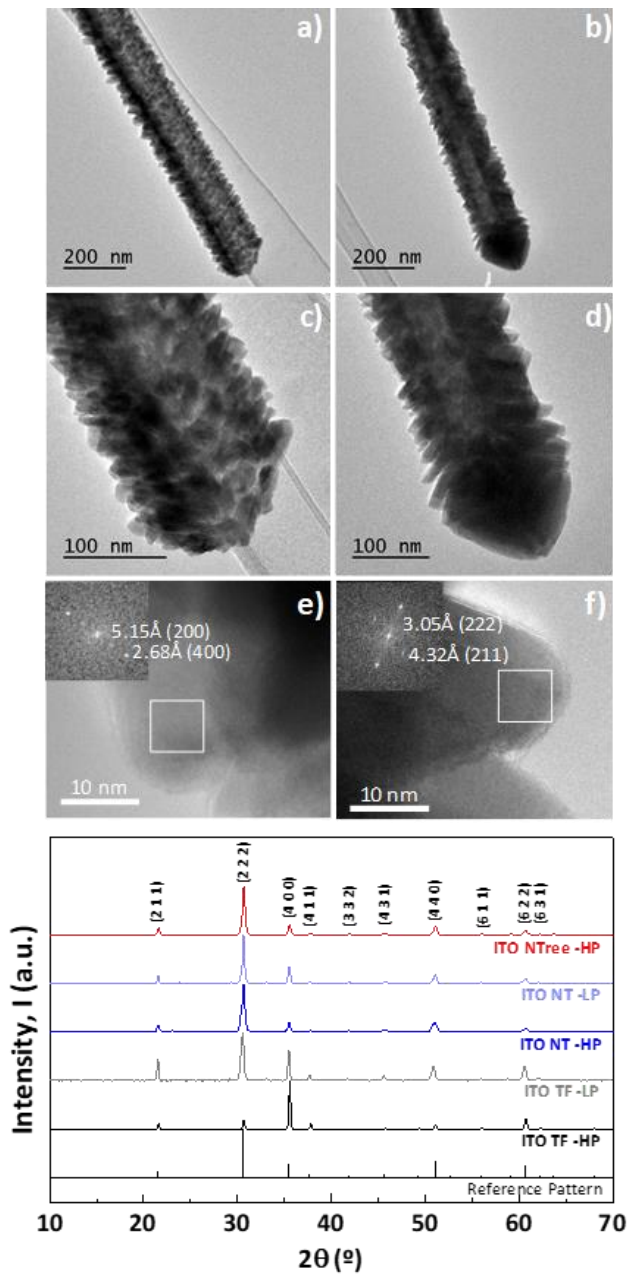

**Figure 3. High crystalline quality of the produced thin films, nanotubes and nanotrees.** TEM (a-d) and HRTEM (e-f) micrographs of the ITO nanotubes samples prepared under LP (left-side) and HP (right-side) conditions. g) XRD indexed diffractograms comparing the supported nanostructures and the reference thin films.

We have also studied the crystalline texture using the Textural parameter, $T_{hkl}$, (see Figure S2). The deposition conditions lead to the formation of highly textured samples, with the preferential growth of planes (211) and (411) for the three types of samples prepared at both LP and HP conditions. These results demonstrate on one hand that this synthetic approach produces highly crystalline nanotubes and, on the other hand, that the thin film microstructure is extended to the 1D and 3D nanostructures but at a reduced characteristic scale. Importantly, both columns and crystal domains are smaller for the nanotubes than for the reference thin films.

To elucidate the surface and "bulk" chemical composition of the ITO nanotubes we carried out X-ray Photoelectron Spectroscopy (XPS) and Rutherford Backscattering Spectrometry (RBS) characterizations. Figure S3 and Table S3 summarize the spectra acquired for the different samples. Photo-electron peaks corresponding to In 4s, 3d, 3p, and 3s; Sn 3d and 3p; O 1s; C 1s; and Auger electron peaks (In MNN, O KLL and C KVV) have been labelled in the spectra. Besides, we quantitatively analysed the elemental composition using the peak zone spectra to obtain the atomic percent composition of the surface layer (Table S3). We obtained that the ratio of Sn to Sn + In is 0.10 in the LP ITO nanotubes and 0.12 in HP ones. We can confirm that the target composition is reproduced in the surface layer of the LP NTs and it is slightly enriched in Sn for the HP conditions. In the RBS spectra in Figure S3, one can observe high-energy (1000-1400 keV) signals that correspond to ions backscattered upon interaction with Sn and In atoms. Furthermore, there are two distinctive spectral shapes in two different regions. For the high energy region (corresponding to the sublayer closest to the sample surface), the amount of counts is lower. This finding is in good agreement with a higher porosity related to the nanotube layer. The lower energy region (related to the layer closer to the substrate) presents a higher amount of counts, corresponding to the seed thin film layer formed at the interface with the substrate. For both the HP and LP conditions, the O to In + Sn ratio exceeds 1.5 at the interface with the substrate and gets closer to 2 at the nanotube layer. Such excess of oxygen might be associated with the presence of carbonaceous species.

**Optical properties.**

As introduced above, the main application of ITO thin films is as a transparent conductive oxide in optoelectronic devices. Figure 4 a-b) presents photographs of the samples at different stages of thin films, NTs, and NTrees fabrication. The as-prepared ITO thin films present high transparency under visual inspection (panel (a) on the left). Figure 4 c) and d) compares the UV-vis-NIR ballistic transmission spectra for ITO thin films deposited at LP (c) and HP (d) conditions. In both cases, the spectrum is divided into three sections. Firstly, one can see that transmittance in the UV region (λ < 400 nm) is low. This is related to the onset position to the absorption across the fundamental optical band gap (see Table 1, estimated by Tauc plot). The optical band gap for bulk ITO is 3.53 eV [46], in our case, 3.54 eV and 3.42 eV, for the reference thin films deposited under LP and HP conditions, respectively. This difference might be related to the composition of the samples slightly enriched in Sn for the HP conditions in comparison to the LP case. Secondly, in the visible (VIS) range, both samples present the transmittance spectra featuring the interference fringes characteristic of transparent thin films. The refractive index and extinction coefficient wavelength dependence of these thin films were obtained by fitting these transmittance spectra with WVASE software using a Lorentz complex refractive index model with three oscillators (Table 1, see also Experimental).

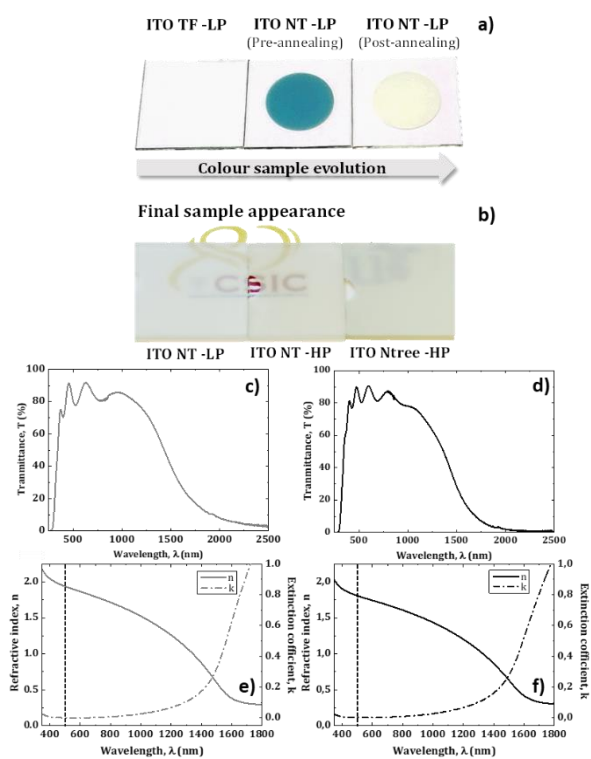

**Figure 4. From transparent thin films to strong light scattering in 1D and 3D nanoarchitectures.** a-b) Photographs of the synthesized samples as labelled. Ballistic transmittance spectra and estimated refractive index and extinction coefficient c) and e) for LP and d) and f) for HP thin films.

**Table 1.** UV-vis-NIR spectra analysis results of ITO thin film samples (λ=500 nm)[a]

| Sample/ Measure | Transmittance, T ±0.5% | Refractive index, n | Extinction coefficient, k | Bandgap, BG (eV) |
|---|---|---|---|---|
| ITO_TF_LP | 80 | 1.94 | $1.6 \cdot 10^{-3}$ | 3.54 |
| ITO_TF_HP | 85 | 1.81 | $2.8 \cdot 10^{-3}$ | 3.42 |

[a] All the fittings we carried out have an adjustment $R^2$ value >0.99.

These values are in good agreement with previous reports [47–51]. The extinction coefficients in the visible range (Figure 4 e) and f)) are very low for both samples, which is a desirable characteristic for the intended application of these films. It is worth stressing that the difference between the refractive indices ($n_{LP} > n_{HP}$) may be related to the higher porosity in the HP samples than in the LP samples, which is likely related to the more pronounced columnar growth (see Figure 2). Finally, for wavelength longer than 1000 nm, both spectra show an abrupt drop in the transmittance in the NIR range. This is related to the conductivity mechanism of ITO and, consequently, to the free-carrier absorption requiring the excitation by an IR photon of an electron at the bottom of the conduction band towards higher energy positions within the same band [1]. Once the ONWs are formed, the sample becomes intense blue (a)-middle) corresponding to the $H_2$-phtalocyanine Q-Band absorption in the range between 600 and 700 nm[42] and remains blueish after the deposition of the ITO shell. The post-annealing treatment of the samples (see Experimental section) completely removes the organic template leading to the white colour of the samples (Figure 4 a) and b)). In a good agreement, the UV-VIS-NIR spectra in Figure S4 comparing nanotube samples deposited under LP and HP conditions are dominated by scattering effects in the visible range that reduce their transparency. Besides, the transmittance spectra show the expected drop in the NIR range corresponding to highly conductive ITO nanomaterials.[47–51] Such a whitish dispersive colour of the nanostructured samples is important for applications of these nanoelectrodes in photoelectrochemical and light-emitting devices.

To explore the utility of our nanostructures in this context, we carried out an additional optical characterization of the NTs and NTrees samples prepared under high-pressure conditions making use of an integrating sphere (LP Ntrees are included herein for the sake of comparison, Figure S5 gathers the corresponding results for LP thin films and nanotubes). Figure 5 presents total and diffuse transmittances ($T_T$, solid line, $T_D$, dashed line) (a, b), total and diffuse reflectances ($R_T$, solid line, $R_D$, dashed line) (c, d), and the absorptance (e, f), i.e. the ratio between the light absorbed by the material to intensity of the incident light, for the layers prepared under HP conditions. The diffuse spectra take into account light scattered in all directions except for the specular and ballistic effects, and thus, represent an integrated measure of a single or multiple scattering events. The difference between the total and diffuse components results in the direct value, i.e. specular reflectance and ballistic transmittance, meaning that the light propagation remains along the same straight line as the incident light. In this case, for NTs and Ntrees, we cannot estimate the optical band gap because the absorption edge in the 400 – 500 nm range is already dominated by strong scattering effects. However, it is worth noting that the transmission edge for the NTrees samples is red-shifted with respect to the nanotubes and thin film (Figure 5 a)), this effect is also appreciable in the absorptances presented in panel e). We hypothesize that this observation might be related to the reduction of the bandgap in the nanotrees samples.

Concerning the transmittance, the spectrum for the thin film depicts a nearly zero value for the diffuse component. In the case of ITO NTs and Ntrees, the curves for the total ($T_T$, solid line) and diffuse ($T_D$, dashed line) transmittance overlap at low wavelengths and separate above 390 nm and 560 nm for the nanotubes and nanotrees, respectively. This means that for the shorter wavelengths the direct transmittance is close to zero, thus validating the extremely high light dispersion of the ITO NTs and NTrees. The total transmittance decreases from thin films to nanotubes, and from nanotubes to nanotrees. The highest transmittances of the 1D and 3D nanostructures in the 750 to 1250 nm range.

Looking at the reflectance spectra, we find a similar behaviour for the thin film, i.e. the low reflectivity of this sample corresponds to the specular component, with no contribution from scattered light. For the NTs sample, the diffuse component is the most predominant below 500 nm (dashed and solid lines overlap at low wavelengths). Above this value, the reflectance is still dominated by the diffuse component although there is a significant contribution from the specular reflectance. Strikingly, for the ITO NTrees both total and diffuse reflectance spectra overlap in the VIS and NIR ranges. In the VIS range, the total reflectance increases from the thin film to the nanotubes and from the nanotubes to the nanotrees, with a maximum value of 50 %. It is worth emphasizing that both the nanotubes and nanotrees act as scattering centers for the incident light, behaving as random optical materials. Thus, in the visible region, diffuse light reflection (in different directions from the specular one) dominates. Likewise, a very significant fraction of the transmitted light is also diffuse (in different directions from the ballistic direction).

These results have two promising applications. First, we have developed an extremely efficient diffusor. Figure S6 compares the Haze factors ([diffuse transmission / total transmission] x 100) estimated for the three types of samples in the range between 400 and 800 nm which yields values higher than 80% for the nanotree samples. This is an outstanding value with potential applications,[52–54] for example, in photocatalytic processes such as water splitting where high light scattering increases the light trapping.[55] In the case of light emission, white-colour nanoelectrodes can be used in the white light LED technology where the light-scattering properties of the phosphors are usually utilized as a diffusive medium to obtain uniform emission without hot spots or angular colour distribution.[56] Second, these scattering layers can be implemented as electrodes in photovoltaic and light-emitting devices. The latter application demands higher intensities of the emitted light by enhancing the external quantum efficiency (EQE). Hence, the use of scattering sheets (external or internal) is common, for example, in devices such as OLEDs to improve the outcoupling efficiency of light generated in the device. In our case, the scattering shell would at the same time serve as an electrode to improve light outcoupling efficiency, as it has been previously reported for metal networks.[57] The next step is to study the reflectance and transmittance spectra for the 1D and 3D nanostructures in the NIR range. As explained above, the low transmittance of the samples in this range is related to the conductivity mechanism in doped wide-bandgap metal oxides. Therefore, in good agreement with the Drude equivalent model,[58] the ITO thin film transmittance drops below a certain value becoming almost zero for wavelengths above 2000 nm (Figure 5 a-b)). On the contrary, the reflectance increases for the NIR range reaching up to 60 % for long wavelengths (Figure 5 c-d)). The trend is similar for the transmittance spectra of the nanotubes and nanotrees with a much sharper drop in the case of the 3D nanostructures that present a total transmittance below 2% for wavelengths longer than 1500 nm and nearly zero above 1750 nm (Figure 5 b)). However, the reflectance of the 1D and 3D ITO nanostructures decreases in the NIR. Consequently, these samples show extremely low values of reflectance in this range, namely below 1% for the nanotrees (Figure 5 d)) in the range from 1250 to 2500 nm, thus acting as broadband antireflective coatings. This characteristic is more pronounced for the NTrees and NTs produced under HP conditions (see Figure S5 for comparison), thus, the absorptances shown in Figure 5 e-f) for the HP samples increase dramatically as the wavelength increases in the NIR region of the spectra. The 1D samples feature a plateau of 90-92% for wavelengths higher than 1800 nm and the 3D samples feature a plateau at even higher levels of 99-100% for wavelengths above 1600 nm. In both cases, especially in the latter case, the ITO samples perform like perfect light absorbers in the infrared, inhibiting the transmission and reflection of electromagnetic waves. The perfect absorption (PA) of these ITO supported nanostructures might be attributed to the near-zero values of the refractive index in the NIR region, due to the near-zero values of the electric permittivity in this spectral region.[59] Importantly, the ITO presents a characteristic epsilon-near-zero (ENZ) point in the NIR whose position is a function of the doping-level.[60] Several authors have reported PA in ITO thin films on metallic substrates,[61,62] combined with plasmonic gratings.[63] Yoon et al. reported a broad-band PA in the range 1450-1750 nm by stacking ITO films with different doping levels under attenuated total reflection conditions. Other authors have demonstrated electrically controlled absorption modulation of ITO metafilms in the ENZ region.[63,64] However, very different from the previous reports, our supported nanotubes and nanotrees structures deposited undder HP conditions show a flat and broad-band complete absorptance in the range 1600-2500 nm and do not show multiple PA maxima. Such a behaviour most likely implies that there are different levels of doping in the nanostructures composing the ITO-NT and ITO-NTrees layers. However, the relationships between the level of doping, the nanostructure (nanotubes and nanotrees), and their specific spatial distribution (nanotubes and nanotrees thicknesses and length, density, and branching degree) determining the ENZ properties of the nanostructured ITO and potential electrical tuneability need further studies in order to fully understand the broadband PA observed. Previous attempts to develop broadband near-perfect absorption surfaces relied on the use of metamaterials, surface plasmons, or resonant optical cavities.[65–68] Common shortcomings of these efforts are narrow spectral bands and complex fabrication procedures with limited scalability for large-area applications[29] Recently, several authors have reported alternative solutions based on the formation of one-dimensional nanomaterials such as carbon nanotubes, Si nanoarrays, Cu nanowires, and ITO nanowires.[29,68-70] In our case, the characteristics of the supported nanotubes and nanotrees allow the straightforward use of these nanomaterials in a lithography-free process and on a broad variety of substrates without additional processing, patterning, or special optical arrangements.

**Electrical characterization of layers and single NTs.**

Figure 6 and Table 2 summarize the most important findings regarding the electrical characterization of the samples. The aim is to elucidate the two key electrical resistances, one corresponding to the layer of NTs and NTrees supported on the ITO decorated substrates and the other one corresponding to the individual nanotubes. Figure 6 a) shows the I-V curves obtained for the five types of samples, including the LP and HP thin films determined by a four-probe macroscopic platform (see Experimental Section). All these curves present an ohmic linear relation between the current (I) and the voltage (V) in the studied range (-0.5 and 0.5 V) with different slopes corresponding to the variations in resistance (R) of the samples.

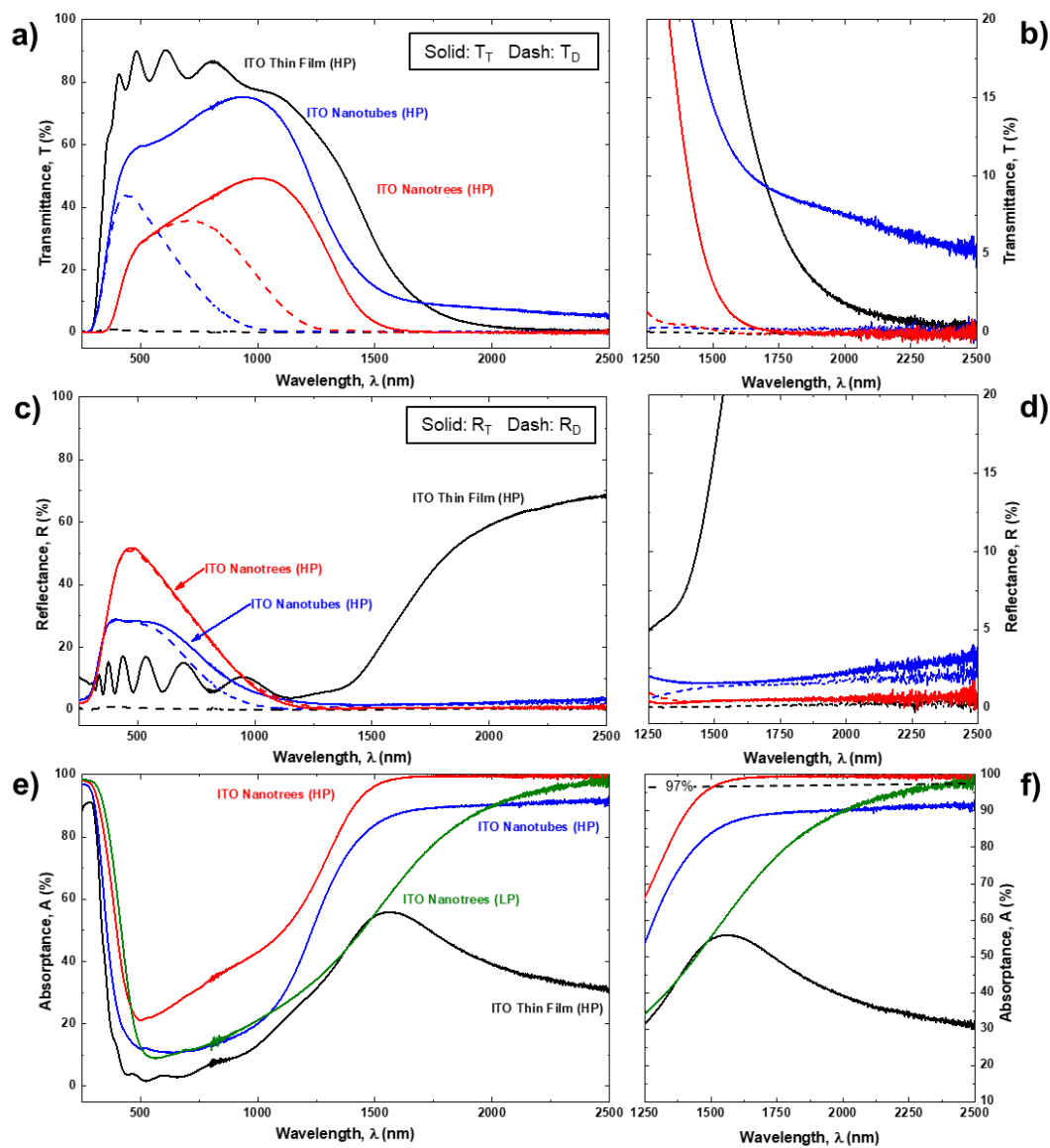

**Figure 5. Nanotubes and nanotrees working as Random Optical Materials in the VIS and Ultra Broadband Perfect Absorbers in the NIR.** UV-vis-NIR spectra of ITO HP samples: a-b) total and diffusive transmittance (TT, TD) and c-d) reflectance (RT, RD). b) and d) are the zoom-in in the NIR range corresponding to a) and b) respectively; (e) absorptance. The corresponding spectrum for LP NTrees is included for the sake of comparison.

Table 2 gathers the estimated R values with the lowest limits corresponding to the thin film samples. Sheet resistance and resistivity of these samples are competitive with the ITO films deposited by RF magnetron sputtering.[47–51] Besides, the sheet resistance and resistivity of the ITO thin film prepared under LP are almost two times lower compared to the HP sample. The macroscopic resistances of the nanotubes samples follow the same trend, being in both cases higher than for the thin film counterparts though still presenting quite low values. Impresively, the resistance for the LP Ntrees is lower than for the 1D counterpart. This improvement in the lateral connectivity can be directly linked to the formation of long secondary branches (see Figure 2) which connect adjacent ITO trunks. Such characteristic is not as pronounced under the HP conditions producing shorter but numerous secondary nanotubes showing high contact resistance from one tree to the other. It must be stressed at this point that the estimation of sheet resistance and resistivity is not straightforward for the 1D and 3D nanostructured samples because of the difficulty inherent to the precise estimation of the conducting cross section area. Therefore, we note that the comparison between the resistances in those cases is only semi-quantitative. Thus, the accurate elucidation of the NTs electrical conductivity requires the characterization in a single-wire approach, i.e. contacting the NTs as individual items.

For the advanced single-wire characterization, the NTs were removed from the substrate and deposited flat on Si (100) wafers as detailed in the experimental section. Figure 6 b) shows a characteristic SEM image taken after the FIB deposition of four equivalent Pt electrodes with areas around 3 x 3 μm². Nanotubes with similar diameters and lengths are contacted aiming for a fair comparison. In the second step, the ITO nanotubes are characterized by a 4 nanoprobe platform from Kleindiek Nanotechnik installed in the SEM (Figure 6 d)). I-V curves in Figure 6 c) are representative of both types of NTs. Here we note that the corresponding values in Table 2 were calculated after analysis of three different NTs from each condition and five I-V curves. In all the cases, linear current versus voltage curves were observed in all measurements. In good agreement with the previous results, the nanotubes resistivity for the HP conditions is higher than for the LP. The resistivity for both samples is in a very low range, and in the LP case, it is comparable to the reported values for single-crystalline ITO NWs fabricated by VLS (2.4 · 10$^{-4}$ Ω·cm)[24] (included in panel c) as a reference value). It is important to stress the differences of our structures with respect to the 1D and 3D single-crystal ITO nanostructures reported in the literature, mostly formed by the catalyst driven mechanisms.[22–24] First of all, although we produce polycrystalline nanotubes, the results of the single-wire electrical characterization support that the high crystalline quality and good connectivity between ITO grains forming the NTs shells allow an effective carrier transport along the NT length. Besides, in the above examples, all the nanotubes are domed, however, the same procedure can be modified to generate open-ended nanotubes by simply increasing the temperature rate slope during the evacuation of the organic template (see additional information elsewhere).[45] Secondly, the thickness and length of the nanotubes are adaptable and controllable by tuning the ONWs length and the ITO shell deposition time.

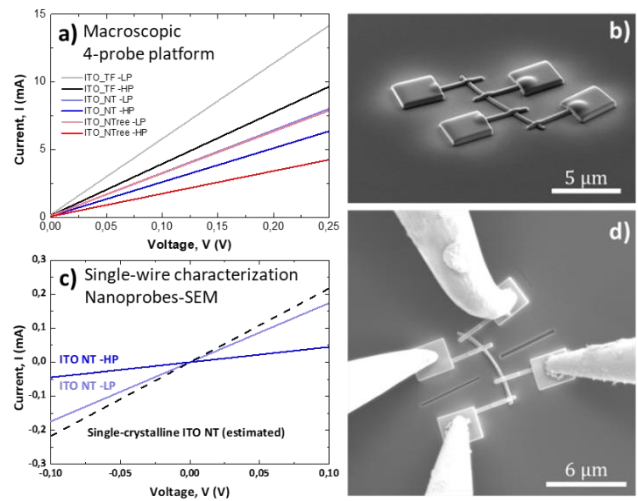

**Figure 6. Competitive conductivity and resistivity of the highly textured thin films and nanotubes.** a) 4-probe electrical characterization of ITO thin film (LP and HP), nanotubes (LP and HP) and nanotrees (HP) I-V macroscopic measurements and c) I-V curves comparative between ITO nanotubes (LP and HP) nanoscopic measurements, the estimated curve for a single-crystalline ITO nanowire has been included as a reference.[24] b) Representative SEM images of the electrical nanoscopic characterization by deposition of four Pt pads by FIB and d) in-situ SEM microprobe measurements.

**Table 2.** Estimation of the resistance, sheet resistance, and resistivity by macroscopic and nanoscopic 4-point probe characterization.

| Sample/Measure | Resistance, R (Ω) / R$_{sheet}$ (Ω/sq) | Resistivity, ρ (Ω·cm) |
|---|---|---|
| ITO_TF -LP | 1.93 ± 0.09 / 16.1 ± 0.8 | (2.48 ± 0.12) · 10$^{-4}$ |
| ITO_TF -HP | 3.29 ± 0.24 / 27.4 ± 2.0 | (4.3 ± 0.4) · 10$^{-4}$ |
| ITO_NT -LP | 4.2 ± 0.2 | - |
| ITO_NT -HP | 5.6 ± 0.4 | - |
| ITO_NTree -LP | 3.7 ± 0.3 | - |
| ITO_NTree -HP | 6.51 ± 0.13 | - |

| Sample/Measure Single wire | Resistance, R (Ω) | Resistivity, ρ (Ω·cm) |
|---|---|---|
| ITO_NT -LP | 480 ± 110 | (3.5 ± 0.9) · 10$^{-4}$ |
| ITO_NT -HP | 180 ± 40 | (15 ± 7) · 10$^{-4}$ |

As an additional advantage over VLS, VS, and other template alternatives, the formation of nanotrees is straightforward by repeating the deposition protocol as many times as desired. Moreover, the growth does not require any modification of the template or the generation of defects on the ITO single crystal surface. Finally, the conformal character of the ITO shell and formation of ITO layer in the free area among the nanotubes may improve the lateral connectivity of the 1D and 3D nanostructures in comparison to the single-crystal counterparts.

## Experimental

### Substrates

We have fabricated ITO thin films, 1D, and 3D nanostructures supported on different substrates attending to the characterization technique requirements. In general, for each experiment a set including the following substrates was covered: Glass slides, fused silica slides, (100) single crystalline P-type doped silicon wafers, and (100) single crystalline intrinsic silicon wafers.

### Synthesis

ITO thin films (ITO_TF) were deposited on Si (100), fused silica under two conditions. The deposition method was magnetron sputtering at two deposition conditions addressed as low pressure, LP, and high pressure, HP. The pressure was regulated by a butterfly valve and controlled by a Pirani pressure meter, the Ar fluxes by a mass flow controller, and the deposition rate and thicknesses by a quartz microbalance located at a close position from the substrates. Cylindric magnetron with a 3'' ITO disc target (SnO$_2$ 10wt.% doped In$_2$O$_3$) (Kurt J. Lesker Company) excited under radiofrequency (13,5 MHz pulses) and argon plasma. Low Pressure conditions (LP) refers to 5·10$^{-3}$ mbar, Ar flux 30 cm$^3$ min$^{-1}$, source power 75W, and High Pressure (HP) to 2·10$^{-2}$ mbar, Ar flux 45 cm$^3$ min$^{-1}$, source power 150W. The substrate holder temperature was controlled at 350ºC with a spinning velocity of 20 rpm.

Secondly, ITO nanotubes (ITO_NTs) were formed on the surface of the different substrates following the fabrication method consists of four steps (see Scheme S1): i) formation of nucleation centers, ii) growth of supported organic nanowires, iii) formation of ITO shell and iv) evacuation of the organic core from the 1D hybrid core@shell nanostructures to leave an empty ITO nanotube. Step i) The features on the surface of the substrate act as NW growth nucleation centres, in this work, the seed or nucleation layer is an ITO thin film deposited by magnetron sputtering under LP conditions to increase the surface roughness. Substrate temperature for the deposition of the ONWs was 210 ºC meanwhile for the formation of the ITO shell, the temperature was increased up to 350ºC. Finally, All the samples were annealed at 350 ºC in Ar atmosphere to achieve steady-state conditions.

All the samples were annealed at 350 ºC in Ar atmosphere to enhance their optical transparency.

**Characterization**

Three different SEM instruments were utilized: microscope S4800 from Hitachi for the overall microstructural characterization; microscope GeminiSEM 300 from Zeiss utilized during the 4-probe electrical characterization (see below) and, FEI Helios Nanolab 650 for the ITO NTs assembly for single-wire nanoelectrode characterization. The micrographs were treated with ImageJ free available software to carry out measurements and statistical analysis of different magnitudes such as film thickness, nanotubes diameter, and length.

For the TEM characterization, NTs were removed from the substrates by scratching with a diamond tip and then deposited in a holey carbon Cu grid. Bright field and High Resolution (HREM) electron microscopy images were acquired in a JEOL 2100Plus operated at 200kV equipped with a LaB6 filament and a CCD camera (Gatan). The HRTEM micrographs were analyzed with the TIA Reader software generating a digital diffraction pattern to measure interplanar distances and to indexing them. The crystalline structure was analyzed by X-Ray Diffraction (XRD) spectrometer in a Panalytical X'PERT PRO model operating in the θ - 2θ configuration and using the Cu Kα (1.5418 Å) radiation as an excitation source [PANalytical X'Pert HighScore database: Indium Oxide; Reference code: 00-006-0416]. The crystallite size was determined with PANalytical X'Pert HighScore Plus software, which applies the Scherrer equation for the calculations. The texture coefficients or Lotgering factors $T_{hkl}$ was calculated applying the equation from reference.[71] The surface In/Sn ratio was determined by X-ray photoelectron spectroscopy (XPS) PHOIBOS 100 hemispheric multichannel analyzer from SPECS using the Al Kα radiation as an excitation source. General and peak spectra were acquired with a pass energy of 50 and 30 eV, respectively. Samples were characterized after the annealing treatment without any additional surface conditioning. The atomic percentage of quantification was estimated using CASA software. Rutherford backscattering spectroscopy (RBS) was performed at the 3 MV Tandem Accelerator of the National Center for Accelerators (Sevilla, Spain) with an alpha beam of 1.6 MeV and a passivated implanted planar silicon detector at a 165° scattering angle. The RBS spectra were analyzed using the SIMNRA software.

The optical characterization was done by UV-vis-NIR spectroscopy (UV-vis-NIR) in the transmittance and reflectance modes. In this technique, a wavelength sweep from 2500 to 200 nm was set to analyze the near-infrared absorption, the visible range behaviour, and the gap presence in the UV region. The transmittance spectrum could be fitted to obtain the dependence of the refraction and extinction coefficients on radiation wavelength, which was implemented by applying a Lorentz Model of three oscillators with the optical analysis software by J.A. Woollam (WVASE®).[72] Two UV-vis-NIR spectrophotometers were utilized, a PerkinElmer Lambda 750 UV/vis/NIR model for specular transmittance and reflectance and an Agilent Technologies Cary 5000 Uv-Vis-NIR equipped with an integrating sphere to acquire absorptance A, diffuse and total transmittance ($T_D$ and $T_T$), diffuse and total reflectance ($R_D$ and $R_T$).

The electrical characterization was carried out at two different levels aiming at both, macro- (nanotubes forming supported layers) and nano- (individual nanotubes) scales. In both cases, the 4-point probe method was utilized. Details on the estimation of resistivity and sheet resistance are gathered at the Supplementary Material Section S7. The source meter was a Keithley 2635A system with a resolution in the range of 100 pA. The measurements were performed for the thin film and nanotube layer samples deposited on fused silica substrates. The equivalent characterization for individual nanotubes requires additional processing of the samples to contact the NTs by microscopic electrodes. This step enables the four nanoprobe electrical measurements assisted by SEM. For this purpose, the ITO NTs are removed from the substrate by scratching with a diamond tip and pressing the scratched area sample against a Si (100) intrinsic piece (note that these pieces present a native $SiO_2$ layer acting as an electrical insulator).

Focused Ion Beam (FIB) technique is used to form Pt patches and pathways with the controlled size and thickness contacting the NT in four different points along its length. The equipment used was an FEI Helios Nanolab 650. The experimental FIB conditions: 30 kV and 40 pA. Contacts dimensions: 4 x 0.14 x 0.50 µm (wire) and 2.5 x 3 x 0.3 µm (square). Precursor: Methyl Cyclopentadienyl trimethyl Platinum (($CH_3$)$_3$Pt(CpCH$_3$)). After the nanostructured electrode fabrication, the samples were characterized in the Zeiss Gemini SEM 300 equipped with four Kleindiek manipulators and nanoprobes (see Figure S7).

**Conclusions**

We have synthesized ITO thin films by magnetron sputtering operating in low- and high-pressure conditions (LP and HP, respectively) with a transparency in the visible range above 80% and low resistivity ($\rho < 5 \cdot 10^{-4}$ Ω·cm). These values are very competitive with the previous reports based on RF and DC sputtering deposition. We have synthesized crystalline ITO nanotubes and nanotrees with crystal sizes lower than 170 nm, homogeneously distributed along the nanotube cavity, and highly textured. In parallel, we have revealed the advantages of the combination of the one-reactor method and the use of ONWs as 1D soft-templates supported on processable and scalable substrates. XPS, XRD, and UV-VIS-NIR results are in good agreement with the formation of highly conductive nanostructures with high absorbance in the NIR. The estimated values for sheet resistance and resistivity indicate that these nanostructured electrodes meet the expectations for their implementation in light-emitting devices and photoelectrochemical processes. Besides, we have demonstrated the potentiality of 1D and 3D nanostructures as optical random media in the VIS range and ultra-broadband perfect absorbers in the range 1600-2500 nm. Meanwhile, the LP conditions produce layers with higher refractive index, the optical properties of the HP NTs and NTrees in the NIR make these latter experimental conditions more interesting within the field of perfect absorbers. On the contrary, the outstanding conductivity revealed under single-wire premises and the reduction of the lateral resistance contact for the LP NTs and NTrees correspondingly, endow the nanomaterials produced under LP with critical features for their implementation in optoelectronic systems. Thus, the results gathered in this work open the way for applications that require effective control over light absorption/emission including renewable energy harvesting, selective electromagnetic absorption, selective thermal emission, and infrared stealth with potential impacts in the defense, renewable energy, and aerospace industries. In all these cases, our new approach can be used for a wide variety of substrates without the need for any structural patterning.

## Author contributions statement

A. Borrás, A. Barranco, J. R. Sanchez-Valencia and K. Ostrikov conceptualization. A. Borrás, A. Barranco, J. R. Sanchez-Valencia funding acquisition. A. Borrás, J. Castillo-Seoane, J. Gil-Rostra writing original draft. All the authors were involved in the investigation, methodology, validation and reviewing and editing of the article.

## Conflicts of interest

There are no conflicts to declare.

## Acknowledgements

We thank the AEI-MICINN (PID2019-110430GB-C21 and PID2019-109603RA-I0), the Consejería de Economía, Conocimiento, Empresas y Universidad de la Junta de Andalucía (PAIDI-2020 through projects US-1263142, ref. AT17-6079, P18-RT-3480), and the EU through cohesion fund and FEDER 2014–2020 programs for financial support. JS-V thanks the University of Seville through the VI PPIT-US and the Ramon y Cajal Spanish National programs. K.O. acknowledges partial support from the Australian Research Council. We thank the following centers for advanced characterization: the Supercomputing and Bioinnovation Center from the University of Malaga, XPS service from ICMS, the CITIUS from the University of Seville, and the High resolution SEM service from University Pablo de Olavide (UPO). GL thanks the funding from the EU H2020 program for the grant agreement 715832 (ERC Starting Grant Nanophom). The project leading to this article has received funding from the EU H2020 program under the grant agreement 851929 (ERC Starting Grant 3DScavengers).

# Supplementary Information

## One-reactor vacuum and plasma synthesis of transparent conducting oxide nanotubes and nanotrees: from single wire conductivity to ultra-broadband perfect absorbers in the NIR

Javier Castillo-Seoane,[1,2] Jorge Gil-Rostra,[1]* Victor López-Flores,[1] Gabriel Lozano,[3] F. Javier Ferrer,[4] Juan P. Espinós,[1] Kostya (Ken) Ostrikov,[5,6] Francisco Yubero,[1] Agustín R. González-Elipe,[1] Ángel Barranco,[1] Juan R. Sánchez-Valencia,[1,2] Ana Borrás[1]*

1) Nanotechnology on Surfaces and Plasma Group, Materials Science Institute of Seville, Consejo Superior de Investigaciones Científicas - Universidad de Sevilla, c/ Américo Vespucio 49, 41092, Sevilla, Spain

2) Departamento de Física Atómica, Molecular y Nuclear, Universidad de Sevilla, Avda. Reina Mercedes, E-41012, Seville, Spain

3) Multifunctional Optical Materials Group, Materials Science Institute of Seville, Consejo Superior de Investigaciones Científicas – Universidad de Sevilla (CSIC-US), c/ Américo Vespucio 49, Sevilla, 41092, Spain

4) Centro Nacional de Aceleradores (U. Sevilla, J. Andalucía, CSIC), Av. Tomás Alva Edison 7, Seville, 41092, Spain

5) School of Chemistry and Physics, Queensland University of Technology, Brisbane, QLD 4000, Australia.

6) CSIRO-QUT Joint Sustainable Processes and Devices Laboratory, Lindfield NSW 2070, Australia.

E-mail: jorge.gil@icmse.csic.es, anaisabel.borras@icmse.csic.es

**Schematic S1.** Scheme of the step-by-step formation of the ITO nanotubes (a) and nanotrees (b).

**a) 1D supported nanostructures formation**

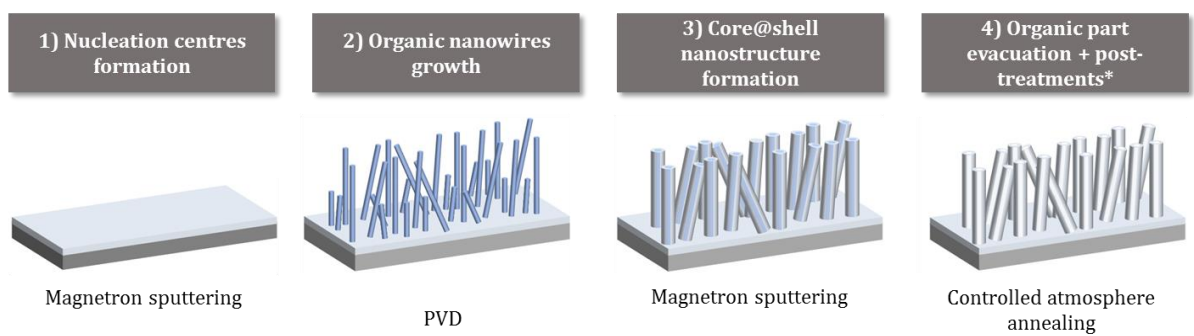

**b) 3D supported nanotrees formation**

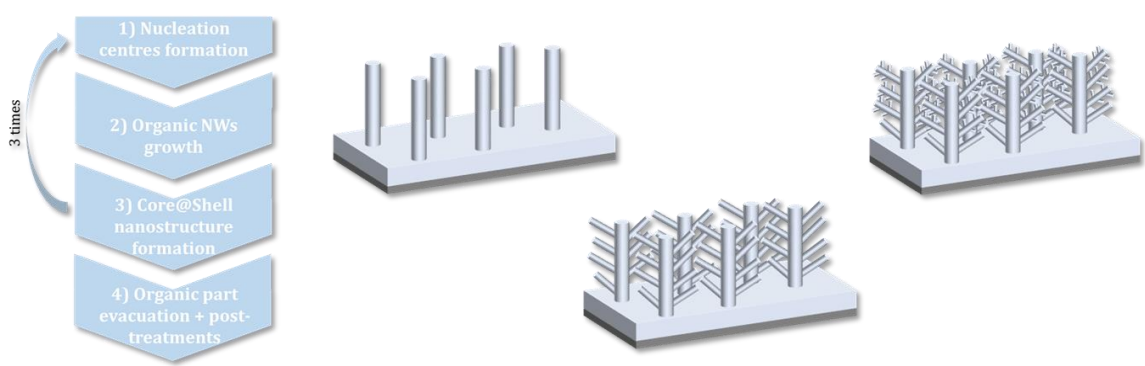

**Step i)** The seed or nucleation layer was deposited under "Low Pressure" conditions with a thickness of 120 nm.

**Step ii)** The fabrication of the single-crystalline ONWs is carried out by PVD in the following conditions: The $H_2PC$ is evaporated in the Knudsen cell starting on its powder (provider details). In this step, the Ar flux was reduced to 10 cm$^3$/min and the substrate holder was cooled to 210 ºC (temperature determined in previous studies to get the best condensation – sublimation ratio of the molecules on the substrate surface that allows a fine nanowires growth),[1] the pressure was set at $1.0 \cdot 10^{-3}$ mbar and the Knudsen cell where was kept the organic compound powder is heated up sequentially until its sublimation temperature (~330 ºC). In order to get a slow deposition rate at the beginning to produce the nanowires nucleation centres, it was established Knudsen cell temperature increments which were 5 ºC in 5 minute intervals, starting at 0.03 Å/s. When the deposition rate became ~25.0 Å/s (microbalance control), it was only necessary to increase the temperature to keep this value approximately constant to get a homogeneous nanowires growth.

**Step iii)** Formation of the ITO shell by magnetron sputtering. The ITO shells were deposited using the same magnetron sputtering configuration and conditions detailed in the previous paragraph, i.e. LP and HP. In this case, there was a variation at the beginning which was the initial substrate holder temperature (210 ºC). This temperature was kept constant in the first deposition period (until a 1 KÅ thickness controlled by microbalance) to avoid the damaging of the organic core acting as template with sublimation temperatures close to 350 ºC. Once the initial layers of the ITO shell are deposited, the substrate temperature is heated up until the according temperature. This same strategy was settled in a previous article[2] to develop TiO$_2$ anatase nanotubes.

**Step iv)** The complete removal of the organic molecules was carried out by annealing in air atmosphere at 350ºC (heating ramp 2ºC/min) for 3 h and the cool down to RT (cooling ramp 5ºC/min). As final proof of concept, we extended the soft template methodology towards the formation of ITO nanotrees (ITO_NTrees) (see Schematic S1 b). The deposition method and the conditions were the same than the ITO_NTs samples but the steps ii) and iii) were repeated three times to each substrate. Using this approach, a nanostructured material with a three-dimensional morphology (a main central nanotube with secondary branching nanotubes) is formed as it is presented in the results section. After the organic core removal, ITO_NTs and ITO_Nanotrees samples were annealed in the main synthesis chamber in an argon atmosphere under a pressure of 2.5·10$^{-3}$ mbar (Ar flux = 30 sccm) for 3 hours at a temperature of 350 ºC.

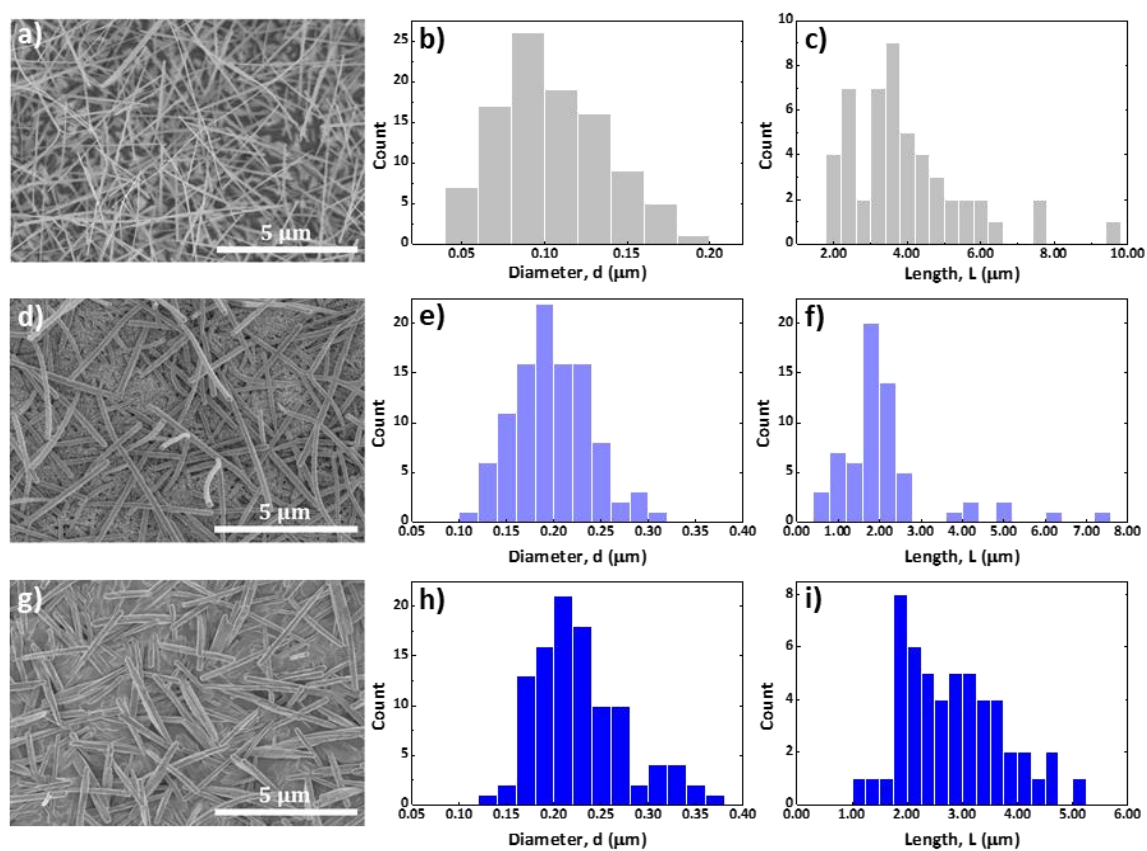

**Figure S1.** Statistical analysis of the ONWs (a, b), LP_NTs (c-d) and HP_NTs (e-f) length and thickness.

**Table S1.** Statistical SEM analysis results of ONWs and ITO nanotubes.

| Sample/Measure | Diameter, D (µm) | Length, L (µm) | Density (n/µm$^2$) |
|:---:|:---:|:---:|:---:|
| ONWs | 0.10 ± 0.03 | 4.0 ± 1.6 | 6.5 ± 0.7 |
| ITO NT -LP | 0.19 ± 0.04 | 2.2 ± 1.2 | 2.7 ± 1.0 |
| ITO NT -HP | 0.22 ± 0.05 | 2.8 ± 0.9 | 3.3 ± 0.6 |

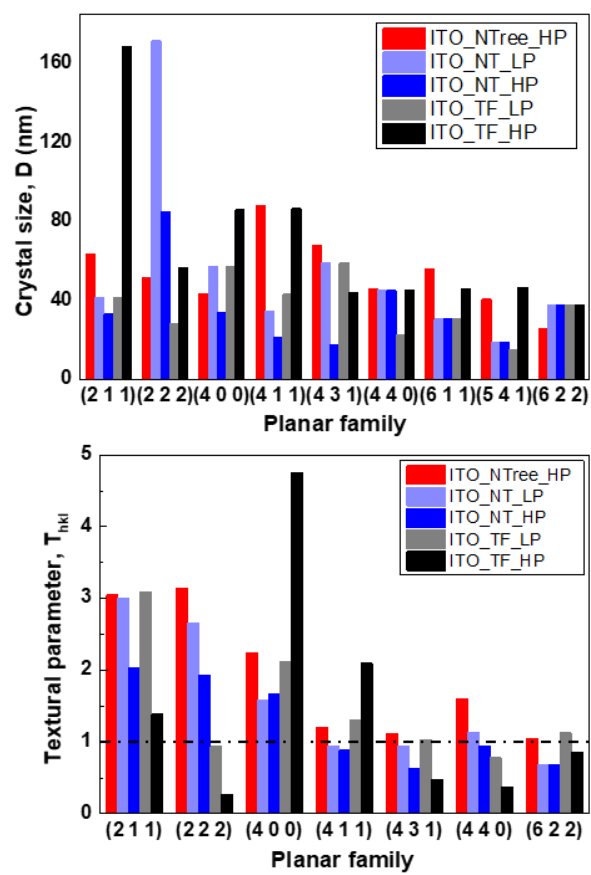

**Figure S2.** Crystal size (top) and textural parameter, $T_{hkl}$ (bottom).

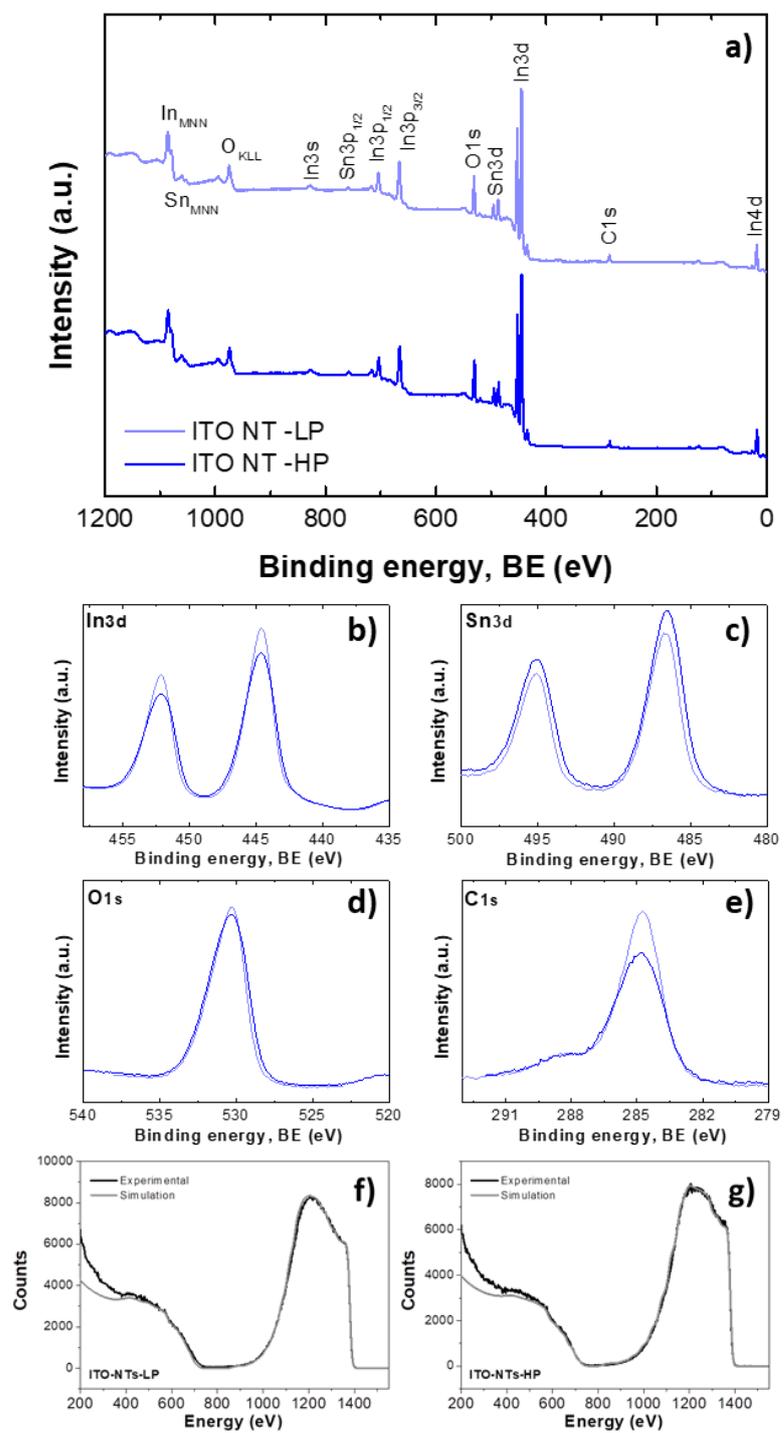

**Figure S3.** General (a) and peak (b-e) XPS spectra for the LP and HP ITO nanotubes; Experimental and simulated RBS spectra for LP (f) and HP (g) ITO nanotubes.

**Table S2.** XPS and RBS quantification results on the ITO nanotubes chemical composition after the post-treatment annealing.

| | Sample/Peaks | | C (1s) (at. %) | O (1s) (at. %) | In (3d$_{5/2}$) (at. %) | Sn (3d$_{5/2}$) (at. %) | Sn:(Sn+In) |
|---|---|---|---|---|---|---|---|
| X P S | ITO_NT_LP | | 19.2 | 41.7 | 35.1 | 4.0 | 0.10 |
| | ITO_NT_HP | | 16.0 | 44.6 | 34.6 | 4.8 | 0.12 |
| | Sample | | Thickness (10$^{15}$at/cm$^2$) | [O] (at.%) | [In] (at.%) | [Sn] (at.%) | O:(In+Sn) |
| R B S | ITO_NT_LP | Substrate ↑ | 649 ± 23 | 61.8 ± 1.6 | 34.8 ± 1.5 | 3.4 ± 0.2 | 1.62 ± 0.11 |
| | | ↓ NTs tips | 680 ± 25 | 66.1 ± 1.5 | 31.0 ± 1.5 | 2.9 ± 0.2 | 1.94 ± 0.13 |
| | ITO_NT_HP | Substrate ↑ | 748 ± 27 | 63.8 ± 1.5 | 33.0 ± 1.5 | 3.2 ± 0.2 | 1.76 ± 0.12 |
| | | ↓ NTs tips | 690 ± 26 | 65.9 ± 1.5 | 31.0 ± 1.5 | 3.0 ± 0.2 | 1.94 ± 0.13 |

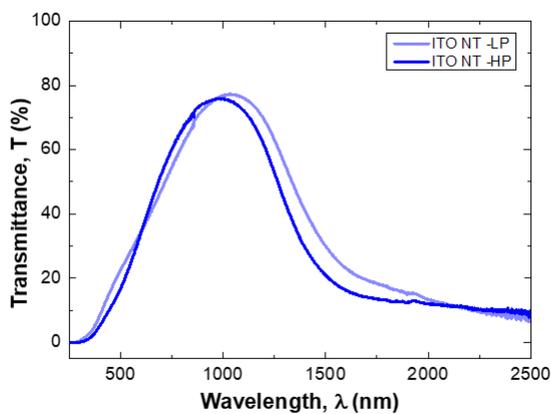

**Figure S4.** Comparison of direct transmittance spectra for the nanotube samples prepared on fused silica under low and high-pressure conditions.

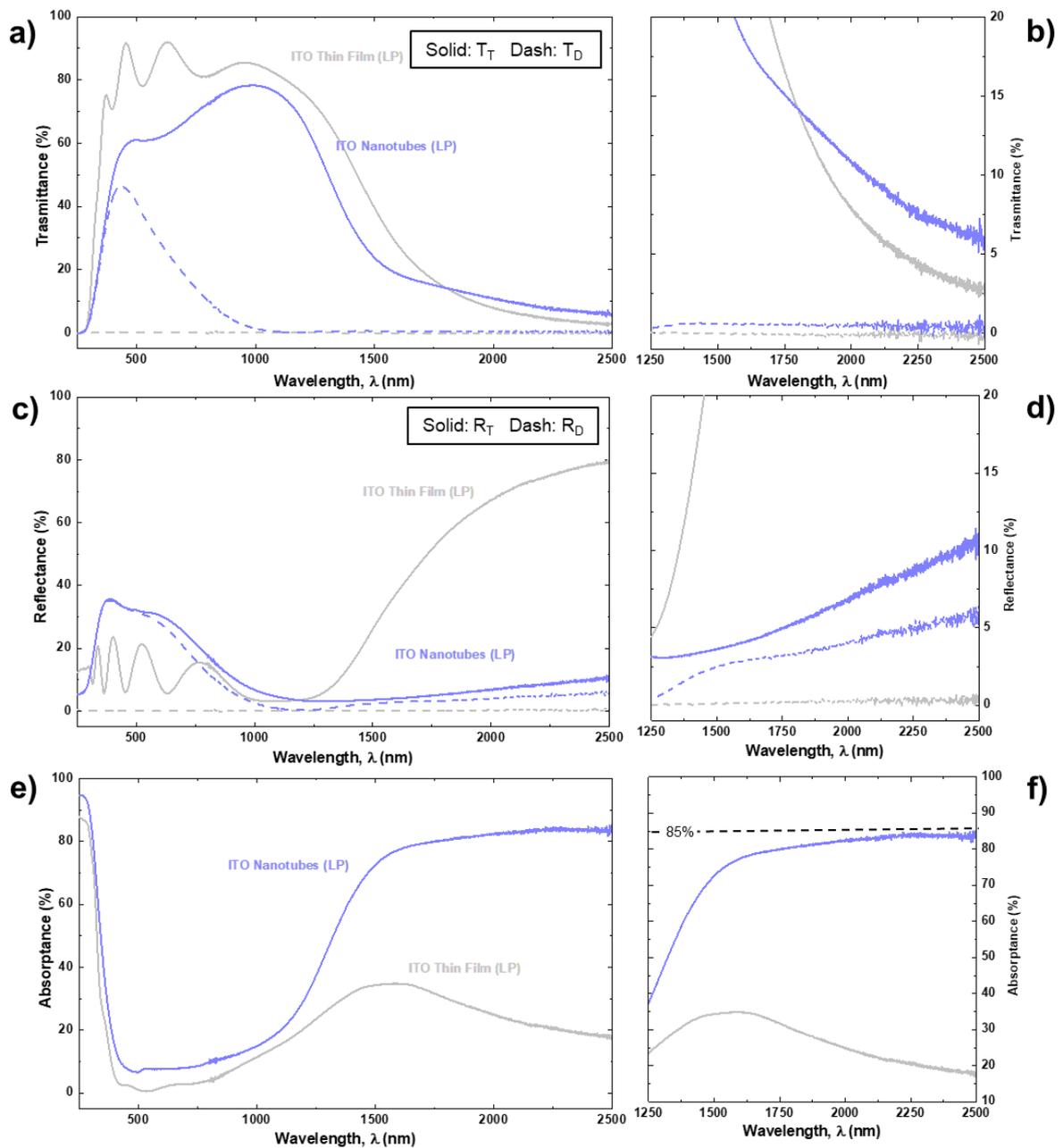

**Figure S5.** UV-vis-NIR spectra of ITO LP samples: (a) total and diffusive transmittance ($T_T$, $T_D$) and zoom-in in the NIR range (b); (c) total and diffusive reflectance ($R_T$, $R_D$) and zoom-in in the NIR range (d); (e) absorptance (e) and zoom-in in the NIR range (f).

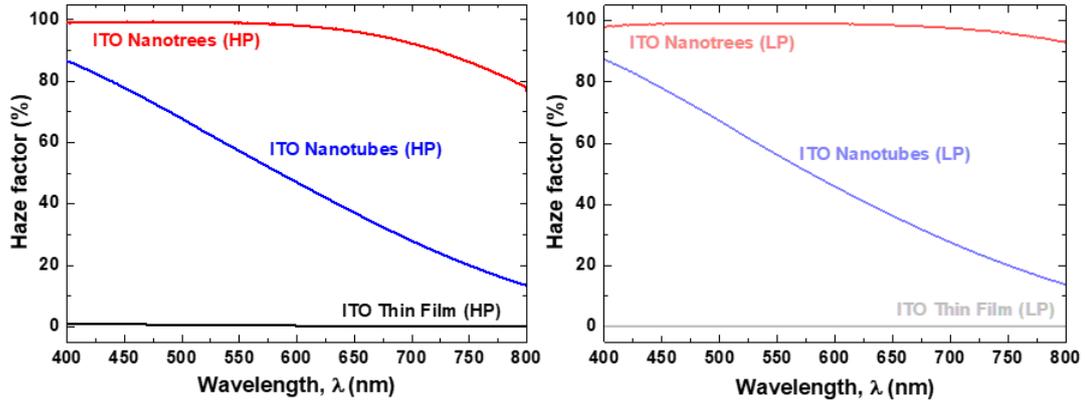

**Figure S6.** Comparison of the haze factors estimated for the three types of samples prepared under HP (left) and LP (right) conditions.

**Section S7.**

The electrical characterization was carried out at two different levels aiming for both, macroscopic, thin films and nanotubes forming supported layers, and nanoscopic, i.e. individual nanotubes, scales. In both cases, the 4-point probe method was utilized. This methodology allows the characterization of thin films and nanostructures avoiding the problems associated with the contact resistance, $R_c$, with the electrodes. Thus, this approach is often applied to circumvent the handicaps related to the electrical characterization of microstructured materials when the resistance in the contacts between wires and electrodes with the specimen is higher than the intrinsic to the specimen. In a typical experiment, the outer contacts are used to scan the current (I) in sweep mode while the voltage (V) induced between the other two probes is acquired.[3]

In the case of ohmic contacts with a conductive material, the obtained I-V curves usually fit to a straight line providing the estimation of the resistance ($R_{sh}$) by equation (ec.1).

$$R_{sh} = \pi\ ln2\ V\ I = \rho\ t \rightarrow \rho = R_{sh} \cdot t\ (ec.1)$$

Where $R_{sh}$ is the sheet resistance, V the voltage, I the intensity, ρ the resistivity and t the thickness of the deposited material. However, to calculate the real sheet resistance of the film is necessary to apply several correction factors to take into account the finite sample dimensions.[3]

$$R_{sh} = F \cdot \pi \cdot ln2 \cdot V \cdot I\ (ec.2)$$

Three different factors for finite isotropic samples are discussed in the literature: $F=F_1 \cdot F_2 \cdot F_3$ (ec.3). F1 considers the effects of the finite thickness of the samples. For very thin samples t/s < 0.1, where t is the thickness of the sample and s is the separation of the electrodes (0.33 mm in our 4-point cell), this factor is one. F2 take into consideration the probes proximity of the sample edge. Our materials were deposited on fused silica substrates of 2.5x2.5 cm$^2$, so the extreme probes are separated at least 0.75 cm from the edge. The correction factor F2 can be considered as the unity too in this case. And the last correction factor, F3, is linked with the sample size. This factor is around 0.8 considering that the sample size is around 8 times bigger than the separation of the electrodes. Thus, the total correction factor is F=0.8. See additional details in reference.[3]

The macroscopic characterization was carried out in a four-probe cell made by the Nanotechnology on Surfaces research group equipped with a sourcemeter working in sweep voltage/current mode under ambient conditions (Figure S7 a)). The sourcemeter was a Keithley 2635A system with resolution in the range of 100 pA. Measurements were done for the thin film and nanotube layer samples deposited on fused silica substrates.

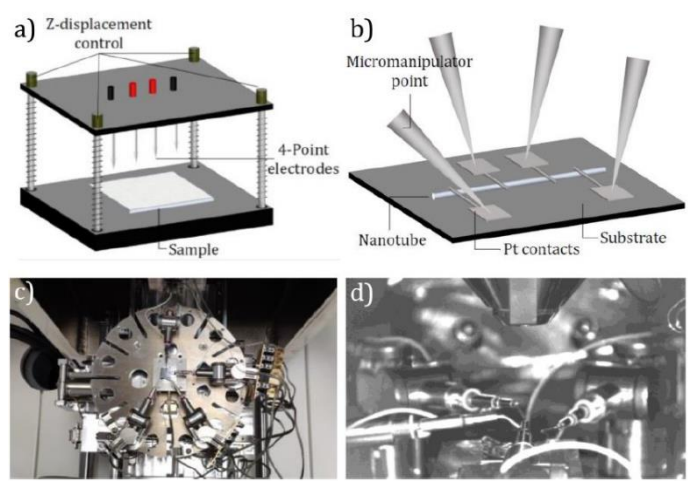

**Figure S7.** Schematics of a) a four-point probe cell and b) four-point probe method applied by micromanipulators on a nanotube; In both configurations, current flows between the two external electrodes while voltage is measured in the two central electrodes. Images of c) the top view of a micromanipulators stage and d) a general view of a micromanipulators stage into a SEM chamber.

The equivalent characterization for individual nanotubes requires additional processing of the samples to contact the NTs by micrometric electrodes. This step enables the four nanoprobe electrical measurements assisted by SEM.

The samples were characterized in a Zeiss Gemini SEM 300 equipped with 4 Kleindiek manipulators and nanoprobes (see Figure S7 c-d). Equation (ec.4) addresses the relationship between the resistance R obtained from the slope[4] of the I-V curves and the resistivity of the NTs characterized by this single-wire approach.

$$R = \rho \cdot L \cdot \pi \cdot (r_2^2 - r_1^2) \rightarrow \rho = R \cdot \pi \cdot (r_2^2 - r_1^2) \cdot L \ (ec.4)$$

Where L is the distance between the two central Pt contacts, $r_1$ the internal nanotube radius and $r_2$ the external nanotube radius.